# Low Particulates Nitrogen Purge and Backfill During Prototype HB650 Cryomodule String Assembly*

T. Ring, M. Quinlan and G. Wu†, Fermilab, Batavia, USA

## Abstract

A low particulate vacuum and purging system was developed to support PIP-II cryomodule string assembly. The overpressure can be controlled at a precision of 1 mbar above the atmospheric pressure regardless of the cavity or string assembly air volume. The system minimized the risk of uncontrolled nitrogen flow during the string assembly. Design features are presented.

## INTRODUCTION

The PIP-II linac consists of several types of cryomodules made of HWR, SSR1, SSR2, LB650, and HB650 cavities [1]. A 6-cavity string assembly is shown in Figure 1. The beamline space volume of PIP-II cavities is considerably higher than a typical 8-cavity 1.3 GHz cryomodule. A string assembly requires the disassembly of cavity beamline flanges when the cavity beamline pressure is under a positive 50 mbar above the atmospheric pressure. This practice was used to purge the cavity beam pipe when the cavity end flanges were removed. This would prevent the particulates from migrating into the cavity.

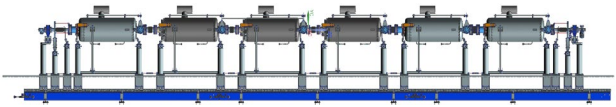

Figure 1: PIP-II 650 MHz string assembly sitting on cleanroom tooling.

During the LCLS-II production [2], it was found that some cavity flange removal did not result in a slow pressure release when the cavity flange was loosened. Instead, a sudden pressure burst was seen. A potential pressure wave could draw the particulates back into the cavity. Unfortunately, it was difficult to reliably repeat this phenomenon to help measure the pressure wave and assess the risk of particulate migration. Nevertheless, this situation was considered an uncontrolled pressure change.

A much larger volume for the PIP-II string assembly means this potential risk could be much worse than that in the LCLS-II string assembly. A standard 50 millibar overpressure would cause high-speed flow when released during cavity flange removal. The mass flow increases when cavity volume increases. 650 MHz and spoke cavities have a substantial risk of uncontrolled pressure change. PIP-II SSR1 prototype string assembly chose to implement this prototype low particulate purging system to avoid the potential pressure wave that could have moved the particulates in the string assembly [3].

* Work supported by Fermi Research Alliance, LLC, under Contract No. DE-AC02-07CH11359 with the U.S. Department of Energy, Office of Science, Office of High Energy Physics
† genfa@fnal.gov

A much lower overpressure, such as 5 mbar instead of 50 mbar, would reduce this risk significantly. A precise pressure measurement, controllable overpressure, and fast detection of pressure drops are required for a new system.

## DESIGN OF THE SYSTEM

### System configuration

A brand new backfill and purging system was designed and being built for cavity coupler installation and string assembly, shown in Figure 2. Nine channels allow the system to build a string assembly with up to eight cavities, including end-group sub-assemblies of a string assembly.

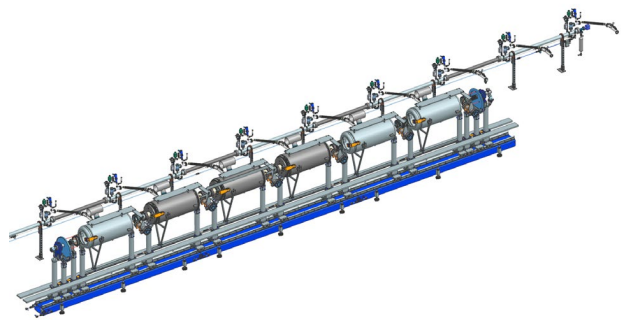

Figure 2: A 9-channel vacuum/purge system in a cleanroom.

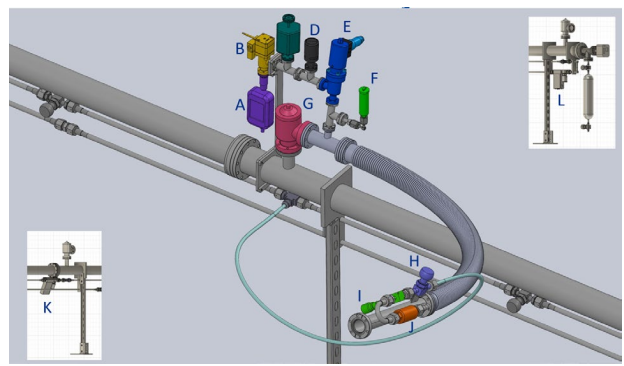

Figure 3: One channel of the purging line with the purging head assembly connected to the main vacuum line.

Two mass flow controllers are located at two ends that can precisely control the flow rate from 0.25 to 15 L/m. The system allows for a closed loop with feedback control or an open-vented system if needed, like the SSR1 prototype string used. The purging entry point is right at the cavity right angle valve (RAV). The venting flow goes through the vacuum hose before the vacuum line isolation RAV in Figure 3.

In Figure 3, the main components are listed as follows:

| | |
|---|---|
| A | Flow meter – Vent |
| B | Electric RAV for opening vent |
| C | Differential pressure transducer |
| D | 1 psi blow-off safety valve |
| E | Manual RAV to isolate vent purge system |
| F | Absolute pressure sensor – UHV rated |
| G | RAV vacuum system isolation |
| H | ¼ turn valve |
| I | Manual flow control needle valve |
| J | Filter |
| K | Upstream MFC with shut off |
| L | Downstream MFC with shut off |

### Software

A LabVIEW® program was developed to log the processing data and control the backfill and purging flow. The diagram in Figure 4 illustrates the control logic.

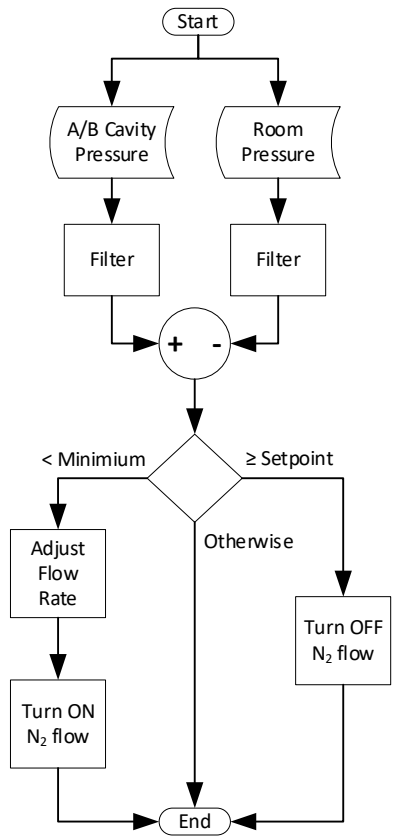

Figure 4: A diagram of control implemented in the software.

Two channels can be controlled simultaneously to allow the counter flow of the cavity sub-assemblies during the cavity connections during the string assembly, as shown in Figure 5. The software also controls the flow rate as needed when the pressure is near the vacuum and maintains the positive pressure.

Controller hardware used a computer with three single boards and National Instruments I/O hardware for a fault-tolerant control system.

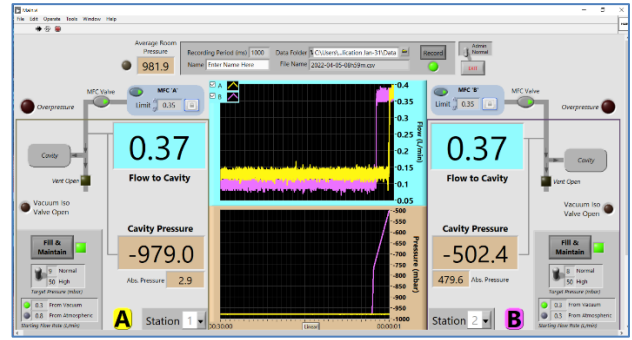

Figure 5: A screenshot of the control program.

## RESULTS AND LESSONS LEARNED

### Results

Two demo channels were built to demonstrate the full system capability, as shown in Figure 6, and tested on an STC test preparation with a high-power coupler. The cavity test showed no degraded field emission and preliminarily validated the system design.

A high flow rate of up to 12 L/m was achieved, while the pressure control could achieve 1 mbar accuracy. The overpressure setpoint was tested up to 100 mbar over atmospheric with one second response time.

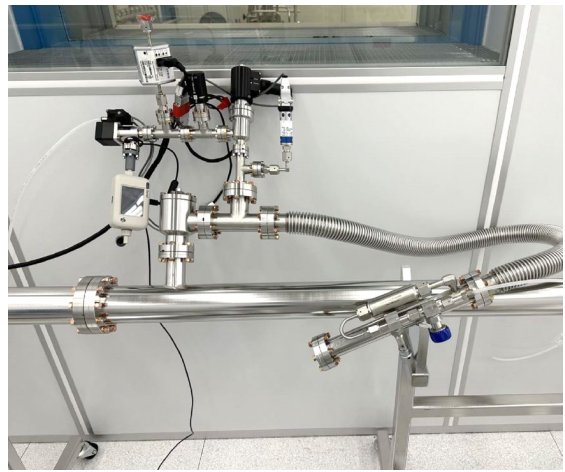

Figure 6: a prototype channel of the purging/backfill.

The two-channel prototype system was later used to complete the HB650 string assembly.

### Lessons learned

During the HB650 string assembly, it was discovered that the pressure reading in the different locations within the clean room varied more than 5 mbar. The prototype 2-channel system was limited to room pressure reading at a single location. The pressure calibration was not a required step and wasn't tested fully. When one channel moved to the downstream end of the string assembly, the pressure reading "fooled" the control program that a cavity was above 5 mbar overpressure relative to the atmospheric pressure. The program stopped supplying the purging flow during the coupler installation. The program continued until the sub-assembly was completed, resulting in a non-ideal assembly step during the string assembly, which could be one of the causes that affected the cavity performance.

The corrective action includes three mitigations. A thorough training of the program will be needed. A pressure calibration would be included before the flow control operation. A pressure reading must be close to the cavity the channel uses to supply the purging flow.

## CONCLUSION

The new purging and venting system avoids the risk of uncontrolled pressure release during cavity flange removal. The over-pressure amount and flow rate can be controlled in real-time.

The system was preliminarily validated through the power coupler assembly.

The HB650 string assembly showed an improvement opportunity, which will be implemented before the next string assembly.

## ACKNOWLEDGMENTS

This manuscript has been authored by Fermi Research Alliance, LLC, under Contract No. DE-AC02-07CH11359 with the U.S. Department of Energy, Office of Science, Office of High Energy Physics.